\documentclass[sigconf,screen]{acmart} 

\usepackage{tabularx}
\usepackage{booktabs}
\usepackage{enumitem}
\usepackage{framed}
\usepackage{graphicx}
\usepackage{placeins}
\usepackage{subcaption}
\usepackage{multicol}
\usepackage{arydshln}
\usepackage{url}
\usepackage{capt-of}
\usepackage{balance}

\newcommand\projectone{\#1}
\newcommand\projecttwo{\#2}
\newcommand\projectthree{\#3}
\newcommand\projectfour{\#4}
\newcommand\projectfive{\#5}
\newcommand\projectsix{\#6}
\theoremstyle{definition}
\newtheorem{observation}{Observation}

\makeatletter
\def\thmhead@plain#1#2#3{%
  \thmname{#1}\thmnumber{\@ifnotempty{#1}{ }\@upn{#2:}}%
  \thmnote{ {\the\thm@notefont#3}}}
\let\thmhead\thmhead@plain%
\makeatother

\makeatletter
\def\th@plain{%
  \thm@notefont{}
  \itshape
}
\def\th@definition{%
  \thm@notefont{}
  \normalfont
}
\makeatother

\setcopyright{rightsretained}
\acmPrice{}
\acmDOI{10.1145/3468264.3473927}
\acmYear{2021}
\copyrightyear{2021}
\acmSubmissionID{fse21ind-p69-p}
\acmISBN{978-1-4503-8562-6/21/08}
\acmConference[ESEC/FSE '21]{Proceedings of the 29th ACM Joint European Software Engineering Conference and Symposium on the Foundations of Software Engineering}{August 23--28, 2021}{Athens, Greece}
\acmBooktitle{Proceedings of the 29th ACM Joint European Software Engineering Conference and Symposium on the Foundations of Software Engineering (ESEC/FSE '21), August 23--28, 2021, Athens, Greece}

\begin{document}

\title{Data-Driven Extract Method Recommendations: A Study at ING}

\author{David van der Leij}
\email{davidvanderleij@gmail.com}
\affiliation{%
  \institution{Delft University of Technology, ING}
  \country{The Netherlands}
}
\author{Jasper Binda}
\email{Jasper.Binda@ing.com}
\affiliation{%
  \institution{ING}
  \country{The Netherlands}
}
\author{Robbert van Dalen}
\email{Robbert.van.Dalen@ing.com}
\affiliation{%
  \institution{ING}
  \country{The Netherlands}
}
\author{Pieter Vallen}
\email{Pieter.Vallen@ing.com}
\affiliation{%
  \institution{ING}
  \country{The Netherlands}
}
\author{Yaping Luo}
\email{Yaping.Luo@ing.com}
\affiliation{%
  \institution{ING, Eindhoven University of Technology}
  \country{The Netherlands}
}
\author{Maurício Aniche}
\email{M.F.Aniche@tudelft.nl}
\affiliation{%
  \institution{Delft University of Technology}
  \country{The Netherlands}
}


\begin{abstract}
  The sound identification of refactoring opportunities is still an open problem in software engineering. Recent studies have shown the effectiveness of machine learning models in recommending methods that should undergo different refactoring operations. In this work, we experiment with such approaches to identify methods that should undergo an Extract Method refactoring, in the context of ING, a large financial organization. More specifically, we (i) compare the code metrics distributions, which are used as features by the models, between open-source and ING systems, (ii) measure the accuracy of different machine learning models in recommending Extract Method refactorings, (iii) compare the recommendations given by the models with the opinions of ING experts. Our results show that the feature distributions of ING systems and open-source systems are somewhat different, that machine learning models can recommend Extract Method refactorings with high accuracy, and that experts tend to agree with most of the recommendations of the model.

\end{abstract}



\maketitle

\sloppy

\newcommand{\keytakeaway}[1]{\begin{framed} \noindent \textbf{Key takeaway:} #1 \end{framed}}

\section{Introduction}
\label{ch:introduction}

Software projects are ever-evolving due to the advent of new functionality, bug fixes, and performance optimizations. With this evolution, however, comes the problem of code degradation.
As a project evolves, the scope and complexity increase and the original design decisions tend to fade.
Refactoring, as defined by \citeauthor{Fowler1999}, is the process of changing a software system in such a way that does not alter the external behaviour of the code yet improves its internal structure.
This process has been shown to improve code quality~\cite{Leitch2003,Kim2014,Gatrell2015}, and its benefits are perceived by developers~\cite{Kim2014}.

Nonetheless, software refactoring does not come without its challenges. 
Any refactoring operation incurs costs~\cite{Kim2014}, and 
its benefits may not be immediately clear~\cite{Ammerlaan2015}. It is also challenging to identify sound refactoring opportunities, i.e., pieces of code that would undoubtedly benefit from refactoring. 
A wide variety of approaches have been proposed to tackle these challenges, such as rule-based approaches, where refactoring opportunities are detected through static rules (e.g.,~\cite{Fokaefs2011,Tsantalis2011,Tsantalis2009}), search-based approaches, where the problem is modeled as a search problem (e.g.,~\cite{Harman2007,OKeeffe2008,Alizadeh2020}) and, finally, machine-learning approaches, where models learn to predict future refactorings (e.g.,~\cite{Aniche2020}).  
In particular, \citet{Aniche2020} experimented with six different machine learning algorithms.
Authors modelled the problem as binary classification, and used code metrics as features to predict whether a piece of code should be refactored.
When using a Random Forest model, with open-source systems as training and test data, authors achieved a precision rate of over 90\%.

Interested in such results, we set out our goal to replicate the study of \citeauthor{Aniche2020} and explore the effectiveness of their approach within ING\footnote{\url{https://www.ing.com}}, a large financial organization. We set our initial goal to explore recommendations of the Extract Method refactoring, which ING software experts deem to be an important refactoring operation and believe that such recommendations would improve the state of their code base. 

Our work can be divided into three parts:

\begin{itemize}
    \item First, we explore the differences in the distributions of the code metrics, which are used as features of the model, between open-source and ING systems (Section~\ref{ch:data-analysis} of this paper). This goal of this initial parts is to shed some light on how different ING code base is from open source, and whether we should expect a model that is trained on top of thousands of open-source systems to work well at ING.
    
    \item Second, we explore the effectiveness of machine learning models in predicting refactoring operations that have happened at ING (Section~\ref{ch:machine-learning} of this paper). We explore different machine learning models, the feature importances of the best performing models, and how the models behave when trained and tested on different sets of internal projects.
    
    \item Third, we collect the perceptions of five ING experts on the recommendations that are provided by the models by means of a user study (Section~\ref{ch:user-study} of this paper). More specifically, we show a set of recommendations of the model (of methods that should and should not undergo an Extract Method) and ask the opinion of experts who do not know what was the recommendation made by the model.
\end{itemize}

Our results show that (i) the feature distributions of ING systems and open-source systems are somewhat different, (ii) that machine learning models can recommend Extract Method refactorings with high accuracy, that different models may work better in different systems, and (iii) that experts tend to agree with many of the recommendations of the model.  
\section{Related Work}

The software engineering community has been studying software refactoring for a long time and from many different angles. When it comes to identify and suggest refactorings, we observe mainly three techniques in the field: (i) rule-based approaches, (ii) search-based approaches, and (iii) machine learning-based approaches.

\textbf{Rule-based approaches.} Such methods apply rules on code metrics, or other aspects of code, to detect code smells or refactoring opportunities.
For example, \citeauthor{Marinescu2004} proposes a metric-based code smell detection technique~\cite{Marinescu2004}.
Authors use logic rules in combination with code quality metrics to detect code smells.
They report an average accuracy rate of 67\% when analyzing nine different types of smells.

\citet{Silva2014} use a similarity-based approach to detect Extract Method refactoring opportunities.
Authors define a heuristic that scores candidates based on code dependencies.
Using this in combination with selecting only the top recommendation per method, they achieve precision and recall rates of both 0.87.

Another approach was proposed by \citet{Moha2010}.
The authors created a framework named DECOR, which they later implemented in a tool called DETEX.
Their tool uses DSLs in the form of rules which generate smell detection algorithms.
These are then applied to the systems that were used to build these DSLs.

\citet{Tsantalis2009} propose a method to detect Extract Method type refactoring opportunities using code slicing.
In a small case study, they report a developer agreement ratio of 5/9 methods.

\textbf{Search-based approaches.} Such approaches model software engineering problems as search-based problems and solve them using search techniques~\cite{Harman2001}.
\citet{Harman2007} proposed a search-based method that applies to Java code.
Their approach is fit for general purpose and allows for multiple fitness functions to present different Pareto optimal metrics.

\citet{OKeeffe2008} describe CODe-Imp, a search-based approach that allows for automatic improvement of code maintainability.
Authors test four different algorithms in conjunction with their tool and find that multiple-ascent hill climbing is the best-performing algorithm.

\citet{aniche-search-based-remodularization} explores the effectiveness of search-based approaches in improving the modularization of a large-scale system. Authors show that the approach is able to find important Move Class refactorings that reduce coupling and increase the cohesion of the industry partner's codebase. Moreover, the developers of the system under study agreed that such classes were in the wrong modules, although they disagreed with the model's suggestion on the new module to move the class to.

More recently, \citet{Alizadeh2020} propose an interactive method that defines an offline and an online phase.
The offline phase collects refactoring solutions using a genetic algorithm to serve the developer.
In the online phase, the developer ranks these suggestions and these rankings are used in the next iteration of the offline phase to constrain the set of refactoring solutions.

\textbf{Machine learning-based approaches.} Such approaches leverage machine learning algorithms that learn how to predict refactoring operations.

\citet{Fontana2013} use a machine learning-based approach to predict several types of code smells, including class-level smells Large Class and Data Class and method-level smells Long Method and Feature Envy.
They apply several machine learning algorithms including but not limited to SVM's, Random forests, and Na{\"\i}ve Bayes.
Using code metrics as input, they achieve up to an accuracy of 0.990 when predicting Long Method using a Random forest type model.

\citet{ArcelliFontana2016} conducted a study comparing several machine learning techniques for code smells.
The authors evaluated 32 different machine learning algorithms to detect four different types of code smells.
They used code metrics as input for their algorithm.
They found that tree-based and naive Bayes algorithms resulted in the best performance in classifying code smells.

Another study by \citet{Liu2018} shows the application of deep learning to detect the Feature Envy code smell.
As input, the approach uses code metrics and code transformed into vectors.
Then, based on the outcome, the approach predicts the destination of a Move Method refactoring operation.
The paper reports an average f1 score of 52.98\% in detecting feature envy and an accuracy score of 74.94\% in recommending Move Method destinations.

\citet{Yue2018a} combine static analysis and machine learning to recommend Extract Method to software clones.
They report an average f1 score of 83\% when testing within projects.
An average f1 cross-project score of 76\% is reported.

Finally, \citet{Aniche2020} experiments with different machine learning algorithms to recommend different types of refactoring, with Extract Method being one of them. The approach relies on different code, process, and ownership metrics of methods that underwent an Extract Method and methods that did not need an Extract Method. Results show that such models can learn and predict Extract Methods with high accuracy. 
\section{An Empirical Study of Extract Method Refactorings at ING}
\label{ch:data-analysis}

\newcommand\rqDataAnalysisOpenSourceVsIndustry{How does code that underwent an Extract Method refactoring in ING and open-source systems compare in terms of code metrics?}

The goal of this section is to empirically understand the characteristics of methods that underwent an Extract Method refactoring, and methods that did not need such a refactoring. Moreover, we also compare the distributions of features between ING's code and open-source systems (used in the related work), as a way to better understand their similarities and differences. 
To that aim, we answer the following research question:

\vspace{2mm}

\begin{enumerate}[label=\textbf{RQ\arabic*}]
    \item \textbf{\rqDataAnalysisOpenSourceVsIndustry}
\end{enumerate}

\vspace{3mm}

\subsection{Methodology}
\label{sec:data-analysis-methodology}

Similar to the work of \citet{Aniche2020},
we make use of the Git history of the ING projects to build a dataset consisting of code that underwent an Extract Method refactoring and code that did not need an Extract Method refactoring.

We identify instances of methods that underwent an Extract Method via RefactoringMiner version 2.0~\cite{Tsantalis2018,Tsantalis2020}.
RefactoringMiner is a refactoring detection tool that analyses Git commits and detects any refactoring operations that have occurred.
\citeauthor{Tsantalis2020} report to achieve precision and recall rates of 99.8\% and 95.8\% respectively for detecting an Extract Method refactorings~\cite{Tsantalis2018,Tsantalis2020}.

To identify methods that did not need an Extract Method, we use the heuristic proposed by \citet{Aniche2020}.
We classify a method as one that did not need an Extract Method if its class did not undergo any refactoring for \(s\) consecutive commits (i.e., the class was changed \(s\) times without any refactoring operation happening).
Deciding the sensitive parameter \(s\) is still an open problem. If we increase \(s\), the tool gets less sensitive about what to class as a non-refactored class as the class needs not to be refactored for more consecutive commits.
Conversely, if we lower the sensitivity, the collector will require fewer steps and confidence before classing a class as non-refactored. \citet{Aniche2020} reports different accuracies when picking different parameters. 
After some exploration, we opt for \(s = 20\) for ING.

From both refactored and non-refactored methods, we collect a large set of code metrics.
Similarly to \citet{Aniche2020}, metrics are collected before the refactoring had occurred rather than after it has been completed, as we want to investigate the method's state for when it was a candidate for refactoring.

We remove any duplicated data points that may exist in our dataset. Duplicated data is a well-known problem for machine learning models, as they can cause training algorithms to have access to the test set, overfit, and inflate performance metrics.
More details on the adverse effects of duplicated code in machine learning models are further elaborated upon in a paper by \citet{Allamanis2019}.

Given that analysing all the 61 metrics that \citet{Aniche2020} use in their models is a daunting and manually-impossible task, we focused on a subset of metrics. In particular, we focus on three types of metrics considered relevant by the ING experts, complexity, coupling, and cohesion:

\begin{itemize}
    \item \textbf{Complexity metrics:}
        \begin{itemize}[leftmargin=*]
            \item \textbf{Lines of code (LOC) \([0, \infty]\)} A longer class or method might indicate more complexity than a short one.
            \item \textbf{Response for class (RFC) \([0, \infty]\)} The sum of all distinct method calls plus the number of methods in a class/method.
                  A higher value indicates more potential interactions and could indicate a higher complexity.
            \item \textbf{Cyclomatic complexity (WMC for classes, CC for methods) \([0, \infty]\)} Indicates branching complexity.
                  For classes, we use the sum of the cyclomatic complexity of the methods in that class.
            \item \textbf{Quantity of unique words (UW) \([0, \infty]\)}
                  A higher value might indicate more responsibilities or interactions with different domains for a certain class/method.
        \end{itemize}
        
    \item \textbf{Coupling metric:}
    
        \begin{itemize}[leftmargin=*]
            \item \textbf{Coupling between objects (CBO) \([0, \infty]\)} Represents the number of connections to a respecting class/method.        
        \end{itemize}
    
    \item \textbf{Cohesion metrics:}
        \begin{itemize}[leftmargin=*]
                
            \item  \textbf{Tight class cohesion (TCC) \([0, 1]\)} Measures cohesion between visible methods.
                  This is calculated by dividing the number of direct connections between a class by the number of possible connections.
            \item  \textbf{Loose class cohesion (LCC) \([0, 1]\)}
                  The same as TCC, but this metric also takes into account indirect connections.
        \end{itemize}
\end{itemize}

\subsection{Datasets}

\begin{table}
    \centering
        \caption{Number of data points, per ING project.}
    \label{tbl:data-analysis-metadata-individual-projects}
\begin{tabular}{lrr}
\toprule
\textbf{Project} &  \textbf{Underwent} &  \textbf{Did not need} \\
 & \textbf{an Extract Method} &  \textbf{an Extract Method} \\
\midrule
ING \#1      &              58 &              37 \\
ING \#2      &             273 &             450 \\
ING \#3      &             152 &              84 \\
ING \#4      &             135 &             212 \\
ING \#5      &              49 &              46 \\
ING \#6      &              52 &              32 \\
Others       &              200 &              125 \\
\midrule
Total & 919 & 986 \\
\bottomrule
\end{tabular}

\end{table}

We analyze and compare two datasets. 
The first dataset consists of open-source systems (OSS) from GitHub.
This dataset was mined by \citet{Gerling2020}.
The projects were sourced from GHTorrent~\cite{Gousios2013}.
From this dataset, \citeauthor{Gerling2020} selected the top \(100,000\) watched projects, and after removing faulty projects, they were left with \(92,280\) projects to analyze.
This resulted in \(616,088\) Extract Method and \(503,393\) non Extract Method instances.
After removing duplicates, we were left with \(449,949\) Extract Method and \(460,974\) non-Extract Method instances.

The second dataset consists of proprietary code from ING.
This dataset initially contained \(18\) ING projects which were chosen with the help of experts.
The data collection resulted in \(2,083\) Extract Method and \(1,483\) non-Extract Method instances.
After removing duplicates for this dataset, we were left with  \(919\) Extract Method and \(986\) non-Extract Method instances. Table~\ref{tbl:data-analysis-metadata-individual-projects} describes the number of data points per project.

When analyzing individual projects, we only analyze projects that have at least 30 methods that underwent an Extract Method refactoring and 30 methods that did not need an Extract Method.
This restriction ensures a higher level of confidence in the observations of individual projects. 
We identify six projects to analyze on an individual level. Note that we still keep data points of the other software systems for when analyzing all projects together.

\definecolor{amber}{rgb}{1.0, 0.75, 0.0}
\newcommand{\higherArrow}{{\Large\textcolor{green}{\(\uparrow{}\)}}}
\newcommand{\lowerArrow}{{\Large\textcolor{red}{\(\downarrow{}\)}}}
\newcommand{\theSame}{{\Large\textcolor{amber}{\(\approx{}\)}}}

\begin{table}
    \centering
    \caption{
        Differences between ING and open-source feature distributions, per metric.
        \higherArrow{} indicates that the ING values of the metric are higher than the open-source values, \lowerArrow{} indicates that ING values are lower than open-source values, and \theSame{} indicates that values in ING and open-source systems are similar.}
        \label{tbl:data-analysis-industry-vs-oss-summary}
    \begin{tabular}{lcc|cc}
    \toprule
    & \multicolumn{2}{c|}{\textbf{Class-level metrics}} & \multicolumn{2}{c}{\textbf{Method-level metrics}} \\
                        & \textbf{Re-} & \textbf{Not} & \textbf{Re-} & \textbf{Not} \\
        \textbf{Metric}                & \textbf{factored} & \textbf{refactored} & \textbf{factored} & \textbf{refactored} \\
    \midrule
    \multicolumn{5}{l}{\textbf{Complexity metrics}} \\
        LOC    & \lowerArrow{}  & \lowerArrow{}  & \theSame{}      & \lowerArrow{}   \\
        RFC    & \lowerArrow{}  & \lowerArrow{}  & \higherArrow{}  & \lowerArrow{}   \\
        WMC/CC & \lowerArrow{}  & \lowerArrow{}  & \theSame{}      & \lowerArrow{}   \\
        UW     & \lowerArrow{}  & \theSame{}     & \higherArrow{}  & \higherArrow{}  \\
    \hdashline
    \multicolumn{5}{l}{\textbf{Coupling metrics}} \\
        CBO    & \theSame{}     & \higherArrow{} & \theSame{}      & \theSame{}      \\
    \hdashline
    \multicolumn{5}{l}{\textbf{Cohesion metrics}} \\
        TCC    & \higherArrow{} & \higherArrow{} & ---             & ---             \\
        LCC    & \higherArrow{} & \higherArrow{} & ---             & ---             \\
    \bottomrule
    \end{tabular}

\end{table}

\subsection{RQ1: How Does Code That Underwent an Extract Method Refactoring in ING and Open-Source Systems Compare in Terms of Code Metrics?}
\label{sec:data-analysis-oss-vs-industry}

\begin{figure*}
    \begin{subfigure}[htbp]{0.45\linewidth}
        \centering
        \includegraphics[width=0.65\textwidth]{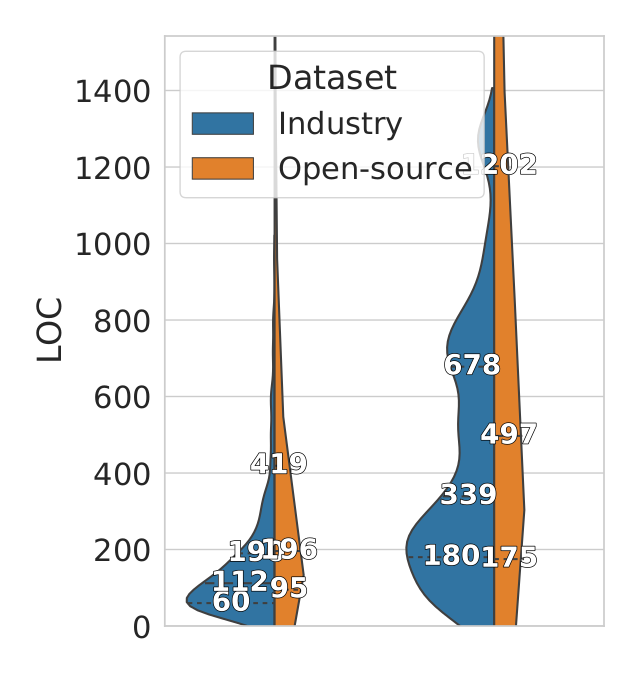}
        \caption{Class-level LOC: The left violin plot indicates classes that contain methods that underwent an Extract Method refactoring.
            The right violin plot indicates classes that do not need to undergo an Extract Method refactoring.}
    \end{subfigure}
    \hfill
    \begin{subfigure}[htbp]{0.45\linewidth}
        \centering
        \includegraphics[width=0.65\textwidth]{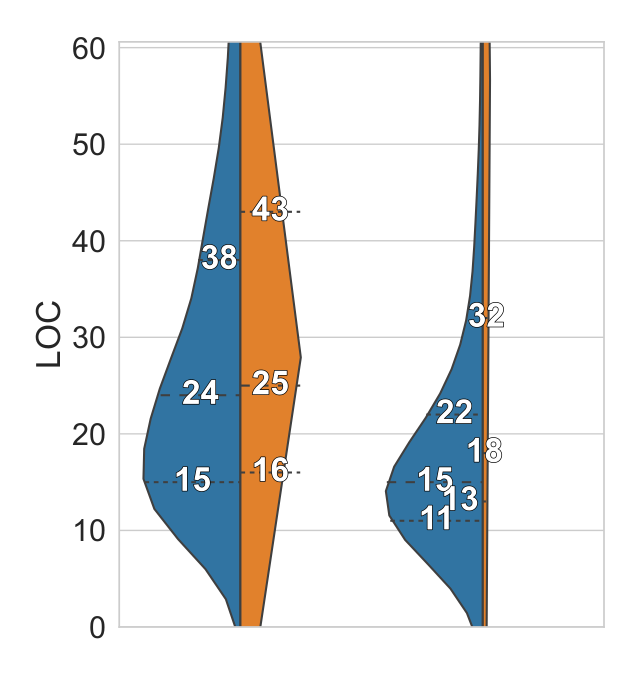}
        \caption{Method-level LOC: The left violin plot indicates methods that underwent an Extract Method refactoring.
            The right violin plot indicates methods that do not need to undergo an Extract Method refactoring.}\label{fig:methodRfc}
    \end{subfigure}
    \caption{LOC distributions for open-source and ING code on both class- and method-level.}
    \label{fig:only-violin-plot-loc}
\end{figure*}

We summarize our findings in Table~\ref{tbl:data-analysis-industry-vs-oss-summary}.
We show an example violin plot of the LOC metric in Figure~\ref{fig:only-violin-plot-loc}.
Due to space constraints, all the other violin plots we used to infer the observations are available only in our appendix~\cite{appendix}. We nevertheless report medians and interquartile ranges near all our observations.

\newcommand{\medrange}[3]{\((MED=#1, IQR=[#2\mbox{--}#3])\)}

\begin{observation}[Classes that contain methods that underwent an Extract Method in ING tend to be smaller and less complex than classes in open-source systems; on the other hand, methods that underwent an Extract Method, in ING and in open-source, are generally similar in terms of size and complexity]
    
We see that ING classes that contain methods that underwent an Extract Method refactoring
are shorter in terms of lines of code \medrange{112}{60}{190} than the open-source classes whose methods underwent the refactoring \medrange{196}{95}{419}.
Similarly, for classes that did not contain methods that underwent an Extract Method refactoring, ING classes are also shorter \medrange{339}{180}{678} as their open-source counterparts \medrange{497}{175}{1202}.

In terms of complexity, we observe that the WMC of classes that contain methods that underwent an Extract Method refactoring in the ING systems is almost half that of its open-source counterpart (\medrange{23}{12}{42} vs \medrange{41}{19}{92}).
We see the same pattern in the class-level metrics of methods that did not need an Extract Method refactoring, where the WMC in ING systems \medrange{57}{22}{142} is again almost half that of open-source \medrange{105}{35}{269}.
We also note that WMC in the open-source classes is much more widely spread than in ING's classes.

At method-level, differences between methods that underwent an Extract Method in ING and in open-source system are generally less pronounced.
ING methods that did not need an Extract Method refactoring \medrange{25}{16}{43} are similar in length to open-source methods that did not need an Extract Method refactoring \medrange{24}{15}{38}.
We also see similar cyclomatic complexity for methods that did not need an Extract Method refactoring in the ING systems \medrange{3}{2}{4} and open-source \medrange{4}{3}{8}.

\end{observation}

\begin{observation}[Coupling is generally similar among classes and methods that underwent an Extract Method in ING and in open-source]

We see that, for methods that underwent an Extract Method, method-level CBO is similar between ING systems \medrange{4}{2}{6} and open-source systems \medrange{4}{3}{6}.
The same holds for methods that did not need an Extract Method, where the median and inter-quartile ranges are equal for both ING and open-source systems\medrange{2}{1}{4} .

On the other hand, the class-level coupling of methods that did not need an Extract Method refactoring i much lower in open-source systems \medrange{20}{9}{41} than in ING systems \medrange{29}{14}{84}.
Most interestingly, however, is the upper quartile, where the open-source has a normal distance from the median, but the CBO of ING systems is significantly high in comparison.
\end{observation}

\begin{observation}[Cohesion is higher in ING classes]

For ING classes that contain methods that underwent an Extract Method refactoring, the inter-quartile range \medrange{0.30}{0.00}{1.00} spans all possible values of the TCC metric.
This is probably due to the high density around 1.00, as seen in the upper part of the violin plot.
This range is much wider as in the same open-source case \medrange{0.16}{0.02}{0.37}.
We see the same pattern, but more pronounced in the LCC metric.

\end{observation}

\begin{observation}[Within the six ING systems, we observe different similarities and differences]

When it comes to the class-level complexity, we observe clusters of similar projects. More specifically, \projecttwo{} and \projectfour{} are somewhat similar to each other, while \projectthree{}, \projectfive{} and \projectsix{} are similar among each other. Interestingly, \projectone{} is different from all others.

We also observe that class-level coupling varies widely in some projects for the class-level metrics of methods that did not need an Extract Method refactoring.
More specifically, the interquartile ranges for all projects but \projectone{} \medrange{16}{12}{22} and \projectthree{} \medrange{23}{22}{29} are very large.
The upper quartiles of some projects, such as \projectfour{} \medrange{30}{19}{320}, are even larger.

On the other hand, at method-level, metrics are very similar between different projects.
We see that for all projects, and for both types of instances, ranges of values do not differ from project to project to a large degree.
For example, if we compare the largest difference between two projects for methods that underwent an Extract Method refactoring, which can be found between projects \projectfour{} \medrange{16}{9}{27} and \projectsix{} \medrange{12}{10}{16}, we see that they are relatively insignificant.
Our appendix~\cite{appendix} contains the plots for the other method-level metrics which show similar patterns.
\end{observation}

\keytakeaway{The differences in metrics between ING and open source systems tend to happen more significantly at class-level, and are less pronounced at method-level. We, therefore, conjecture that a model trained on open-source data will have just reasonable performance.}

\section{The Effectiveness of Machine Learning Models in Recommending Extract Method Refactorings at ING}
\label{ch:machine-learning}

The goal of this section is to explore the effectiveness of machine learning in predicting Extract Method refactorings in the ING code base. Moreover, we explore whether models trained on top of open-source systems are also effective in predicting refactorings in the ING code base, and how much ING models generalise across different ING software systems.

To that aim, we answer the following research questions:

\newcommand\rqMachineLearningIndustryOnIndustry{How effective are supervised machine learning models at predicting Extract Method refactoring opportunities in ING?}
\newcommand\rqMachineLearningOpenSourceOnIndustry{How well do models trained on open-source systems perform in predicting refactoring in ING systems?}
\newcommand\rqMachineLearningIndustryOnIndustryGeneralize{How well do Extract Method models generalise across different ING systems?}

\begin{enumerate}[label=\textbf{RQ\arabic*}]
    \setcounter{enumi}{1}
    \item \textbf{\rqMachineLearningIndustryOnIndustry}
    \item \textbf{\rqMachineLearningOpenSourceOnIndustry}
    \item  \textbf{\rqMachineLearningIndustryOnIndustryGeneralize}
\end{enumerate}

\subsection{Methodology}
\label{sec:machine-learning-methodology}

We follow a similar method to the one proposed by \citet{Aniche2020}.
We model the problem of classifying whether an Extract Method refactoring should be applied as binary classification.
We attach true labels to methods that underwent an Extract Method refactoring,
and false labels to methods that did not need an Extract Method refactoring.
Our feature vectors consist of the collected class- and method-level code metrics. More specifically, we collect 
41 class-level metrics, and 20 method-level metrics. These metrics are extracted using the CK tool\footnote{\url{https://www.github.com/mauricioaniche/ck}} and are listed in our appendix~\cite{appendix}.

We train our models using scikit-learn~\cite{scikit-learn}.
We make use of the same algorithms as used in the paper of \citet{Aniche2020}, except neural networks.
From an exploratory investigation, we found that neural networks did not achieve better performance when compared to our best-performing model.
Because of the large number of hyper-parameters and the long training times, we decide not to investigate this algorithm.
The algorithms used and their corresponding abbreviations are as follows: Random forest (RF), Decision Tree (DT), Logistic Regression (LR), Linear SVM (SVM), and Gaussian Naive Bayes (NB).

In a nutshell, the training pipeline consists of the following steps for each algorithm:
\begin{enumerate}
    \item Pre-process data:
          \begin{enumerate}
              \item Query Extract Method and non Extract Method instances and their corresponding metrics.
              \item Apply the associated labels.
              \item Shuffle the data.
              \item Split the data into a train and test set in a stratified manner.
              \item Scale the features.
              \item Apply feature reduction (LR only).
          \end{enumerate}

    \item Train the model for every hyperparameter combination and investigate their performance.
    \item Record the hyperparameter combination of the model with the highest performance.
    \item Calculate the performance of this model on the test set.
    \item Train a production model with the above parameters using both the training and test set and persist it.
\end{enumerate}

In their paper, \citet{Aniche2020} choose to balance their dataset such that the amount of positive samples is equal to the number of negative samples.
Authors do this because, for most refactoring types, the classes are highly unbalanced, and imbalanced data can lead to problems in machine learning~\cite{He2009}.
In our experiments, we choose to not balance our dataset because of two factors:
First, for the refactoring type Extract Method, this imbalance between classes is not as severe.
We see this by the number of samples in each class displayed in Table~\ref{tbl:data-analysis-metadata-individual-projects}. The ratio is 919 for Extract Method (48\%) to 986 for non-Extract Method (52\%) for industry code and 449,949 for Extract (49\%) to 460,974 for non-Extract Method (51\%) for open-source code.
Second, during exploratory runs of our models, we did not observe significant changes in F1 performance when comparing a model trained on balanced vs imbalanced data.

\subsection{Data Collection, Pre-Processing, and Balancing}
\label{sec:data-pre}

We extract the data from ING systems as described in Section~\ref{sec:data-analysis-methodology}.
In a nutshell, we visit the Git's history of the project and extract instances of methods that underwent an Extract Method and methods that did not need an Extract Method. We then apply a true label to the Extract Method instances and a false label to instances that did not need an Extract Method refactoring. We remove duplicated data points.

We use stratification during the split to ensure classes are not over-represented in the test or training set by chance.
We scale all our features using a MinMaxScaler since this benefits most machine learning algorithms\footnote{Based on sklearn's manual: \url{https://scikit-learn.org/stable/modules/preprocessing.html\#preprocessing-scaler}}.
Finally, we apply feature reduction only for the logistic regression model.

When answering RQ2, we split the entire ING dataset, without any specific separation among projects, into train (80\%) and test (20\%). When analysing the performance of open-source models in industry code (RQ3), we use all industry data as test set. Finally, when analysing the performance of models in unseen ING systems (RQ4), we train the model with the data available for all projects but the one we test on, and we repeat the procedure for all six projects. 

\subsection{Training}

To improve our models' performance, we optimise these hyperparameters by defining a hyperparameter space and exhaustively searching for the combination of parameters that results in the best performance.
We use F1 as our performance metric because we are working with a slightly imbalanced dataset.
We execute sklearn's grid search, which fits a model using all combinations in a pre-defined hyperparameter search space.
For every combination of hyperparameters, we calculate the performance of the resulting model with the help of K-fold validation on the training set where \(k=10\).
Our appendix~\cite{appendix} contains the hyperparameter combinations of the best-performing models.
With the resulting model, we calculate the confusion matrix using the test set.
The raw confusion matrices can be found in the appendix~\cite{appendix}.
These confusion matrices are then used to calculate well-known accuracy, precision, recall, and F1 metrics.

To analyse which features are most important for a model's performance we make use of permutation importance.
This measures the reduction of performance in a particular model on a validation set if we randomly permute a certain feature~\cite{Breiman2001}.
If the performance drops significantly, we know that the feature is important for the model's performance.
Conversely, if it drops only a small amount or not at all we know the feature to be unimportant for the model's performance on that set.
We compute this on the validation set to measure the performance on unseen data.
We permute each feature 50 times to increase our confidence in the performance reduction.

We do not analyse coefficients for linear models (SVM and LR) and the feature importances available in impurity-based models (DT and RF).
We opt for the permutation importance instead since the linear coefficients and impurity-based feature importances illustrate features' importance on the training set rather than its importance for a model to perform well on unseen data.
In addition to this, impurity-based feature importances are biased towards high cardinality numerical features\footnote{Based on sklearn's manual: \url{https://scikit-learn.org/stable/auto_examples/inspection/plot_permutation_importance.html}}.
For reference, plots of the above coefficients and feature importances can be found in our appendix~\cite{appendix}.

Finally, for the best-performing model only, we create a so-called ``production model''.
This model is trained with the previously found best-performing hyperparameter set and uses all data, including the test set.
It is not used for analysis as the test set is used to build it.
This model and its associated scaler are then saved in ONNX\footnote{\url{https://github.com/onnx/onnx}} format.

\subsection{RQ2: How Effective Are Supervised Machine Learning Models at Predicting Extract Method Refactoring Opportunities in ING?}
\label{sec:machine-learning-industry-on-industry}

\begin{table}
    \centering

    \caption{Performance metrics for models trained and tested on ING code.}
    \label{tbl:perf-industry-industry}
    \begin{tabular}{lrrrr}
\toprule
\textbf{Model} &  \textbf{Accuracy} &    \textbf{F1} &  \textbf{Precision} &  \textbf{Recall} \\
\midrule
\textbf{RF        } &     0.934 & 0.935 &      0.899 &   0.973 \\
\textbf{DT        } &     0.850 & 0.843 &      0.855 &   0.832 \\
\textbf{LR        } &     0.824 & 0.825 &      0.794 &   0.859 \\
\textbf{SVM       } &     0.829 & 0.832 &      0.793 &   0.875 \\
\textbf{NB        } &     0.782 & 0.799 &      0.721 &   0.897 \\
\bottomrule
\end{tabular}
\end{table}

We summarise the achieved performance of each model in Table~\ref{tbl:perf-industry-industry}.

\begin{observation}[All the different machine learning models perform relatively well]

From Table~\ref{tbl:perf-industry-industry}, we see that there is no precision rate below 0.721 (Naive Bayes).
For recall, the lowest rate is 0.832 for Decision Trees.
Lastly, if we look at the aggregated metrics of accuracy and F1, we see that they do not drop below 0.782 (NB) and 0.799 (NB), respectively.
\end{observation}

\begin{observation}[Random Forest has the highest performance]

Table~\ref{tbl:perf-industry-industry} shows that Random Forest models achieve the highest values for all performance metrics.
For accuracy, F1, precision and recall we observe values of 0.934, 0.935, 0,899 and 0.973 respectively. We note that Random Forests were also the most accurate model in the work of \citet{Aniche2020}.
\end{observation}

\begin{observation}[Class-level features are most important for model performance]

For all different machine learning algorithms, the top five features that reduce performance the most when permuted are all class-level features. For the Random Forest model, for example, the top five features are quantity of unique words (UW), CBO, LCOM, LOC, and RFC. 
\end{observation}

\begin{observation}[Random Forests are more stable with regard to performance when permuting features in comparison with other types of models]
Out of all models, when we permute features for the Random Forest, the performance drops at most 0.0139.
For all the other models, the maximum drop of 0.0233 occurs for Naive Bayes and happens when permuting CBO. This amount is much much higher as the maximum drop in performance for Random Forest.
We also see performance drops as high as 0.1504 for the SVM model when permuting UniqueWordsQty. This is almost 15 times as high as the maximum performance drop for the Random Forest. 
\end{observation}

\subsection{RQ3: How Well Do Models Trained on Open-Source Systems Perform in Predicting Refactoring in ING Systems?}

\begin{table}
    \centering

    \caption{Performance metrics for models trained on open-source and tested on ING code.}\label{table:machine-learning-performance-oss-on-industry}
    \begin{tabular}{lrrrr}
\toprule
\textbf{Model} &  \textbf{Accuracy} &    \textbf{F1} &  \textbf{Precision} &  \textbf{Recall} \\
\midrule
\textbf{RF        } &     0.687 & 0.748 &      0.611 &   0.963 \\
\textbf{DT        } &     0.606 & 0.664 &      0.564 &   0.808 \\
\textbf{LR        } &     0.761 & 0.794 &      0.678 &   0.958 \\
\textbf{SVM       } &     0.759 & 0.794 &      0.676 &   0.963 \\
\textbf{NB        } &     0.614 & 0.713 &      0.556 &   0.995 \\
\bottomrule
\end{tabular}

\end{table}

We summarise the achieved performance of each model in Table~\ref{table:machine-learning-performance-oss-on-industry}.

\begin{observation}[Open-source-trained models perform reasonably well on ING code, although not as good as models trained on ING's code.]

Table~\ref{table:machine-learning-performance-oss-on-industry} shows that the best-performing model (LR) achieves an accuracy and F1 score of 0.761 and 0.794 respectively.
These are relatively high values, but much lower than the best-performing ING-trained model (RF), which achieves accuracy and F1 scores of 0.934 and 0.935, respectively.

\end{observation}

\begin{observation}[Models trained on open-source code are much better at predicting non-Extract method instances as they are at predicting Extract Method instances in ING code]

Table~\ref{table:machine-learning-performance-oss-on-industry} illustrates that recall scores, which indicate the ability to perform well on non-Extract Method instances, are high in all models, including the best-performing type (LR).
For non-tree type models (SVM, NB, and LR), recall is even higher than their ING-trained counterparts.
The same table shows that precision, a metric that indicates the ability to predict refactoring instances correctly, is much lower for all types of models where the best-performing type (LR) only achieves a score of 0.678 while the best-performing ING-trained model (RF) achieves a precision score of 0.899.

\end{observation}

\subsection{RQ4: How Well Do Extract Method Models Generalise Across Different ING Systems?}
\label{sec:machine-learning-per-project}

\begin{table}
    \centering

    \caption{The average of the performance metrics, when training on all ING projects but one.}\label{tbl:machine-learning-performance-one-out}
\begin{tabular}{lrrrr}
\toprule
\textbf{Model} &  \textbf{Accuracy} &    \textbf{F1} &  \textbf{Precision} &  \textbf{Recall} \\
\midrule
\textbf{DT        } &     0.612 & 0.653 &      0.609 &   0.724 \\
\textbf{NB        } &     0.652 & 0.601 &      0.758 &   0.525 \\
\textbf{SVM       } &     0.711 & 0.671 &      0.762 &   0.641 \\
\textbf{LR        } &     0.720 & 0.692 &      0.787 &   0.668 \\
\textbf{RF        } &     0.744 & 0.771 &      0.715 &   0.860 \\
\bottomrule
\end{tabular}

\end{table}

Table~\ref{tbl:machine-learning-performance-one-out} shows the mean of the performance metrics when training on all but one ING project. We show more detailed numbers in our appendix~\cite{appendix}.

\begin{observation}[Average performance is still reasonable, but much lower than when using the whole dataset, and slightly lower than when training on open-source]

From Table~\ref{tbl:machine-learning-performance-one-out}, we see reasonable aggregated performance metrics, with the best model (RF) having mean accuracy and F1 rates of 0.744 and 0.771
However, we note that performance is much lower than when testing on the whole dataset (RQ2) and slightly lower than the open-source-trained model (RQ3).
\end{observation}

\begin{observation}[For each project, there exists a type of model that performs well, except for project \#1]

For all projects, except project \#1, there is always one type of model that achieves a score of 0.75 or higher.
Project \#1 is an exception, as the highest F1 score is only 0.66 for the RF type model. This is somewhat expected as we observed in RQ1 that Project \#1 has a feature distribution completely different from all other projects.
\end{observation}

\begin{observation}[Different types of models perform well on different projects]

We observe that there is no one type of model that performs the best on one project.
From the F1 score, we see that the RF type model performs best on projects \#1, \#3, \#5, and \#6.
We observe that the SVM type model performs best on project \#4.
Finally, we see that the LR type model performs best on project \#2.
\end{observation}

\keytakeaway{Machine learning models seem to accurately predict Extract Method refactorings in ING systems (accuracy of around 94\%). When models are trained on top of open-source data, the accuracy drops when compared to ING-trained models, but their performance is still high (around 76\%). When trained on ING-systems and tested on unseen ING systems, performance is smaller than when trained with the entire dataset together, but still high (around 74\%). Different models perform better for different ING systems.}

\section{The Perceptions of ING experts on the data-driven recommendations}\label{ch:user-study}

The goal of this section is to compare the perceptions of ING experts with the recommendations provided by the machine learning models.
To that aim, we answer the following research questions:

\newcommand\rqQuantitativeExpert{Do ING experts deem recommended Extract Method refactorings useful/not useful?}
\newcommand\rqQualitativeExpert{Why do ING experts deem recommended Extract Method refactorings useful/not useful?}

\begin{enumerate}[label=\textbf{RQ\arabic*}]
  \setcounter{enumi}{4}
  \item \textbf{\rqQuantitativeExpert}
  \item \textbf{\rqQualitativeExpert}
\end{enumerate}

\subsection{Methodology}
\label{sec:user-study-methodology}

In a nutshell, we show a selection of methods that our best-performing model recommended to (and not to) undergo an Extract Method to the expert. We then ask the expert to decide whether or not s/he would refactor that method, without knowing the prediction of the model.

All predictions served to experts in this section were generated by the best-performing model in Section~\ref{ch:machine-learning}, the Random Forest model that was trained on ING code.
We choose code quality experts at ING and invite them to fill in a questionnaire. The choice was based on convenience and availability.
All participants have substantial experience with programming.
Two participants work with the code displayed in the survey daily, while the other three offer an outside perspective.

The survey consists of 30 questions, each displaying a method originating from ING code.
The set of methods to display were randomly chosen from methods that were added or changed in the ING code base between January 20th 2021 and February 20nd 2021.
The participants do not know whether the model recommended the refactoring or not.

For every method, the ING expert answers two questions.
The first question consists of a scale with four levels where the expert indicates to what extent they find an Extract Method should be applied to the method shown.
A score of 1 indicates that they think it should not be applied at all, while a score of 4 indicates that the operation surely should be applied.
A score of 3 or 4 is interpreted as a sign that the expert thinks an Extract Method should be applied to the method; a score of 1 or 2 is interpreted as a sign that the expert thinks the method should not undergo an Extract Method refactoring.
The second question, an open question, asks the expert to elaborate on their choice.
A concrete example of a question can be found in our appendix~\cite{appendix}.

We settle on a total of 30 methods, as this limits the time spent on the survey by each expert and gives us a reasonable sample size.
The average time for an expert to complete the survey was approximately 42 minutes.
Every expert receives a survey containing the same methods.
Given that we are more interested in evaluating methods for which an Extract Method refactoring is suggested as opposed to when no Extract Method refactoring is recommended, we select 20 methods where the model attached a true label (i.e., the model recommends the refactoring) to the method and ten where it attached a false label (i.e., the model does not recommend the refactoring).

We manually check the answers to the qualitative questions.
We do this by examining and attaching characteristics to each answer.
We then identify patterns in answers and summarize the reasons into characteristics for the quantitative answers.
We only characterize an answer a certain way if the participant explicitly mentions that specific characteristic in their answer.

\subsection{RQ5: Do ING Experts Deem Recommended Extract Method Refactorings Useful/Not Useful?}
\label{sec:surv-quant}

We measure how often the experts agree with the model's prediction to apply Extract Method or to not refactor.
We define the following four situations:
\begin{itemize}
  \item \textbf{Extract Method agreement  (\(R_A\))} The model classifies the method as to be refactored.
        The expert agrees that an Extract Method refactoring is necessary.
  \item \textbf{Extract Method disagreement (\(R_D\))}  The model classifies the method as to be refactored.
        The expert disagrees and thinks an Extract Method refactoring is not necessary.
  \item \textbf{Non-refactor agreement (\(N_A\))} The model classifies the method as to not be refactored.
        The expert agrees that no Extract Method refactoring is necessary.
  \item \textbf{Non-refactor disagreement (\(N_D\))}  The model classifies the method as to be refactored.
        The expert disagrees and thinks an Extract Method refactoring is necessary.
\end{itemize}
With these situations, we define four ratios, similar to accuracy, precision, recall, and F1:

\begin{itemize}
  \item \textbf{Accuracy}: Agreement ratio of the experts with the model's predictions: \(Acc = \frac{RA + NA}{RA + NA + RD + ND}\).
  \item \textbf{Precision}: Agreement ratio of experts for methods where the model predicts Extract Method: \(Precision = \frac{RA}{RA + RD}\).
  \item \textbf{Recall}:  Agreement ratio of experts for methods where the model predicts to not apply an Extract Method refactoring: \(Recall = \frac{RA}{RA + ND}\).
  \item \textbf{F1}: Agreement ratio while taking into account imbalance of classes such as is the case in our experiment: \(F1 = 2 \cdot \frac{Precision \cdot Recall}{Precision + Recall}\).
\end{itemize}
These definitions allow for one-to-one comparison with the theoretical performance.
The raw occurrences and calculations can be found in the appendix~\cite{appendix}.

\begin{table}
  \centering
    \caption{The agreement between ING experts and the recommendation of the models.}
    \label{tbl:user-study-performance-table}
    \begin{tabular}{lrrrr}
        \toprule
        \textbf{Expert  } & \textbf{Accuracy}              & \textbf{F1} & \textbf{Precision} &    \textbf{Recall} \\
        \midrule
        1        & 0.767         & 0.788              & 0.650              & 1.000              \\
        2        & 0.700         & 0.757              & 0.700              & 0.824              \\
        3        & 0.567         & 0.606              & 0.500              & 0.769              \\
        4        & 0.667         & 0.687              & 0.550              & 0.917              \\
        5        & 0.900         & 0.919              & 0.850              & 1.000              \\
        \midrule
        All/Mean & 0.720         & 0.756              & 0.650              & 0.903              \\
        \bottomrule
    \end{tabular}

\end{table}

\begin{observation}[The opinions of ING experts align with the model's predictions to a certain extent]
From Table~\ref{tbl:user-study-performance-table}, we see that the average accuracy and F1 are 0.720 and 0.756, respectively. At an individual level, we see accuracy numbers ranging from 0.567 (the smallest agreement we observed, from expert \#3) up to 0.9 (the most significant agreement we observed, from expert \#5).
\end{observation}

\begin{observation}[Experts agree more with the model when it does not recommend the Extract Method]

Table~\ref{tbl:user-study-performance-table} shows that recall rates are much higher than precision rates, which indicates higher agreement on predictions where the model predicted to not apply Extract Method in comparison with cases where the model did recommend an Extract Method refactoring.
The average recall is 0.903, while the average precision of 0.650 is much lower.
Also, if we look at the minimum precision and recall, we see that the minimum precision of 0.5 is much lower than the recall counterpart of 0.769.
The same is true for the maximum, where the precision of 0.850 is quite a bit lower than the maximum recall of 1.00.
\end{observation}

\begin{observation}[Experts that work with projects where the model was trained on seem to agree with the model's predictions more]

Table~\ref{tbl:user-study-performance-table} shows that for participants \#2, \#5, and who use the analyzed code daily, performance rates are higher on average than for people who do not use the code daily.
The average accuracy, F1, precision, and recall rates are 0.720, 0.756, 0.903 and 0.650, respectively.
Expert \#2's outperforms this, with metrics of 0,700, 0.757, 0.824, and 0.700.
The same is true to an even greater extent for expert \#5's metrics of 0.900, 0.919, 1.000, and 0.850, which is the expert with the highest agreement with the model.
\end{observation}

\subsection{RQ6: Why Do ING Experts Deem Recommended Extract Method Refactorings Useful/Not useful?}
\label{sec:surv-qual}

We observe seven different reasons why experts agree or disagree with the model:
\begin{enumerate}
  \item \textbf{Understandable:} Experts describe the method as understandable and comprehensible enough.
  \item \textbf{Specific:} Experts describe the method as being too domain-specific.
  \item \textbf{Complexity:} Experts mention the (high) complexity of the method.
        This includes matters such as long methods, too many try-catch blocks and, other complexity-related issues.
  \item \textbf{Anti-Patterns:} Experts mention that the method contains an anti-pattern that should be removed. 
  \item \textbf{Potential for refactoring:} Experts mention that, although an Extract Method is not fully crucial, the method would benefit from refactoring.
  \item \textbf{Repetition:} Experts mention the existence of duplicated code in the method.
  \item \textbf{Lack of Readability:} The participant mentions that readability of the code can be improved with the refactoring.
\end{enumerate}

\begin{figure}
  \includegraphics[width=0.45\textwidth]{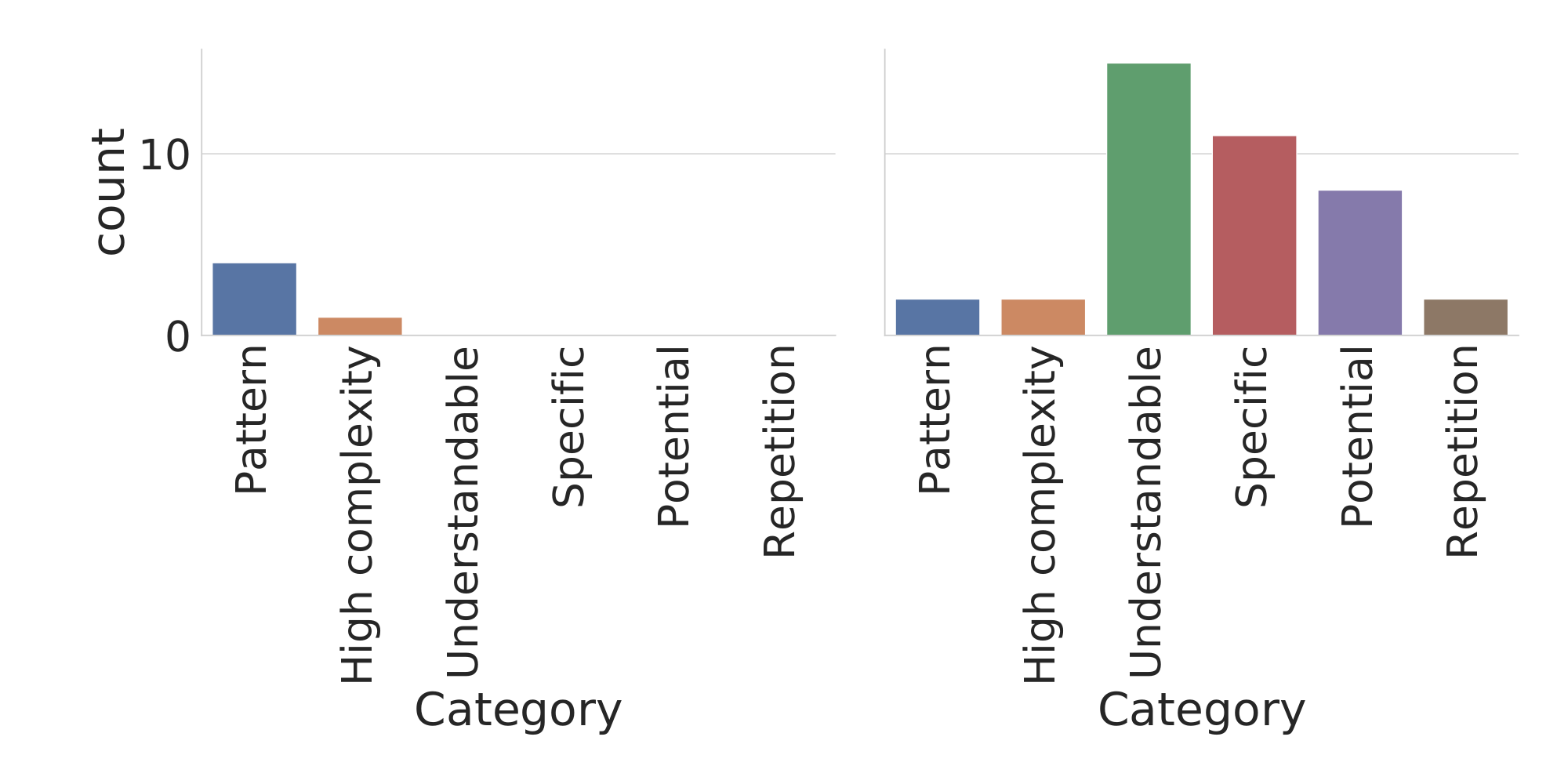}
  \caption[Reasoning when expert disagrees with model]{
    The experts' reasoning for cases where they did not agree with the model's prediction.
    The plot on the left shows the case where the model suggested to not apply an Extract Method refactor to the method and the user did want to apply an Extract Method refactoring.
    The plot on the right shows when the model suggested applying an Extract Method, but the expert did not think the refactoring was necessary.
  }\label{fig:user-study-false-category-distribution}
\end{figure}

\begin{figure}
  \includegraphics[width=0.46\textwidth]{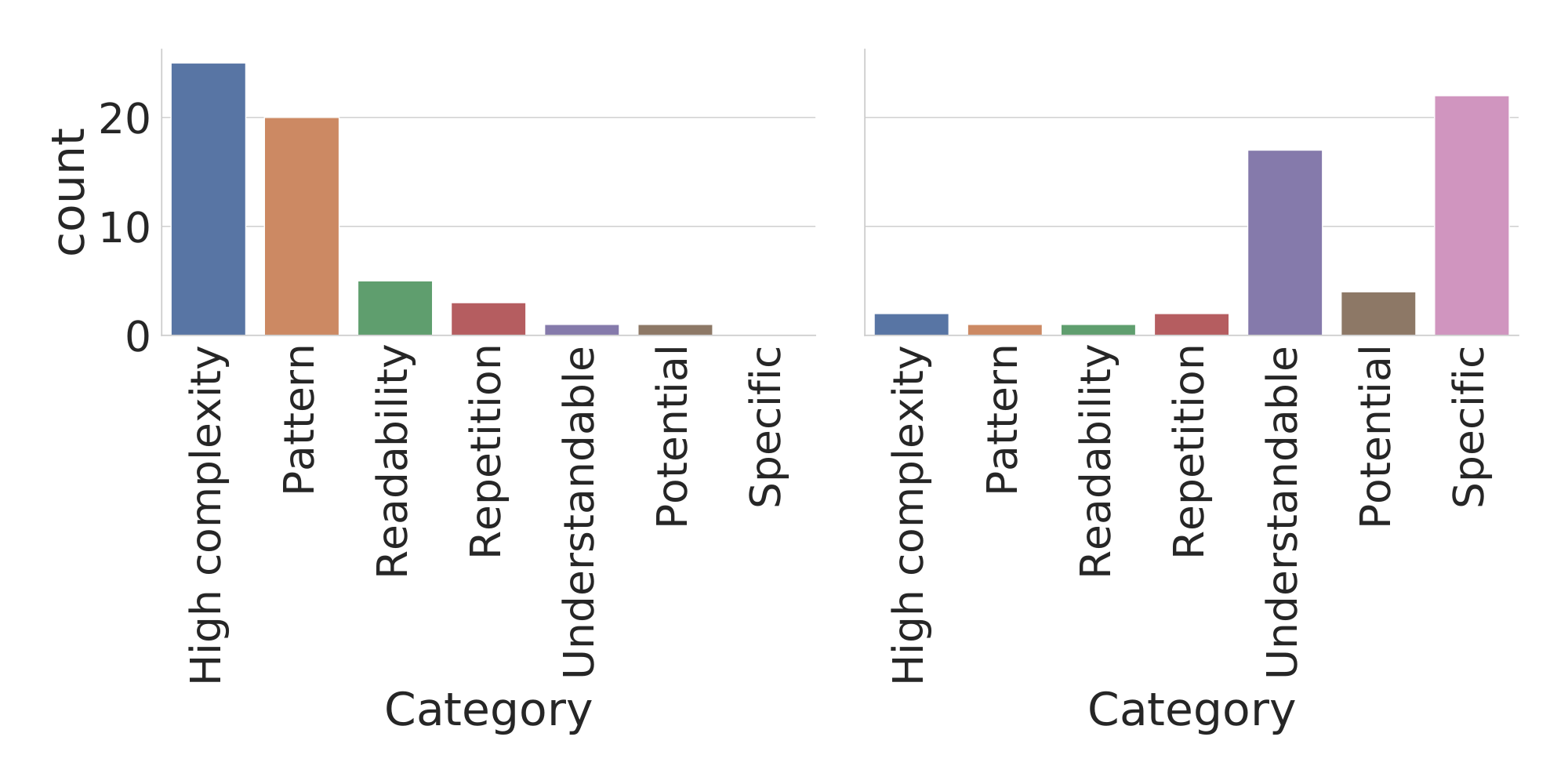}
  \caption[Reasoning when expert agrees with model]{
    The experts' reasoning for cases where they agreed with the model's prediction.
    The plot on the left shows the cases where the model suggested to apply an Extract Method operation on the method.
    The plot on the right shows when the model suggested to not refactor the method.
  }\label{fig:user-study-true-category-distribution}
\end{figure}

We summarize the results of this experiment by plotting the frequency of characteristics in a bar chart.
We present two figures each with two bar charts, one figure displays the frequencies of reasons given where the participants agreed with the model (Figure~\ref{fig:user-study-true-category-distribution}), and in the other plot the frequencies of reasons where the experts did not agree with the model (Figure~\ref{fig:user-study-false-category-distribution}).

\begin{observation}[When experts agree with the model's decision to apply Extract Method, they most often cite the method's high complexity]

We see, from Figure~\ref{fig:user-study-true-category-distribution}, that the most commonly given reason for apply Extract Method to a method (25 occurrences) is that the method is too complex. The second most given reason is that the method contains an anti-pattern or does not adhere to a pattern (21 occurrences).
\end{observation}

\begin{observation}[When experts agree with the model's prediction to not apply an Extract Method, it is because the code is specific enough or already sufficiently understandable]

  We observe from Figure~\ref{fig:user-study-true-category-distribution} that when participants agree with the model's prediction not to refactor, they most commonly mention that the method is is a domain-specific (with 22 occurrences).
  The next most common reason, with 17 appearances, is that the method is understandable enough without further explanation.
  The next reason (the method has ``potential to be refactored'') is much less common but still appears four times.
\end{observation}

\begin{observation}[When experts disagree with the model's prediction to apply Extract Method, they often propose a potential refactoring operation different from ``Extract Method'']
On eight occasions (Figure~\ref{fig:user-study-false-category-distribution}, ``potential''), experts did not agree with the recommendation for an Extract Method refactoring per se, but did mention that the method would benefit from refactoring. Interestingly, while experts mentioned that a refactoring needed to happen in such methods, they were not specific about which one.
\end{observation}

\keytakeaway{The opinions of experts on whether methods should undergo an Extract Method seem to match the recommendations of the model (with an average accuracy of 72\%). Experts that are closer to the software systems agree even more with the models. A common reason for the agreement is that methods are complex; a common reason for disagreement is that the method is already quite understandable.}
\section{Threats to validity}
\label{ch:threats-to-validity}

\subsection{Internal Validity}


\textit{Partially duplicated class-level metrics.}
In this paper, we study Extract Method refactoring recommendations. Given that methods are contained in a class, we train not only on method-level metrics, but also on class-level metrics, as a way to give the classification algorithms some ``context''.
This means, however, that these class metrics are often duplicated in our dataset, as multiple methods exist in one class, and they all share the same class-level metrics.
Such partially duplicated data may inflate performance, which may explain why the performance obtained in RQ1 is much higher than the ones obtained in other RQs.
More research is needed into the role of partially duplicated feature vectors in such models.

\textit{Difference between refactored and non-refactored instances.}
We feed the learners with two types of data points, methods that underwent an Extract Method (the true labels) and methods that did not undergo an Extract Method (the false labels).
While we identify Extract Method instances through the sound strategies implemented by RefactoringMiner, the detection of methods that do not need an Extract Method is the heuristic proposed by Aniche et al~\cite{Aniche2020}.

We used $s=20$ as threshold. This threshold was selected after non-systematic exploration in the ING dataset. Different thresholds might yield different results, and more exploration is needed to define what the best threshold for ING is.

\textit{The selection of experts.}
In our survey, we chose three ING experts that did not know any of studied projects, and two that were part of their development teams.
It could be that the outside perspective was not an accurate assessment of what would be appropriate refactoring opportunities for the relevant code base, since one could argue that doing so would take experience with the code base.
We, however, make no guarantees about whether results would be the same if evaluated by other experts at ING. Future work should expand on the human evaluation of the refactoring recommendations provided by the models.


\subsection{External Validity}

This paper is a case study within a single organization, ING, a large financial organization. It was not our goal to generalize our findings beyond it. That being said, the results we observed in this study are encouraging, and we suggest researchers to replicate this study in other industrial contexts and domains.
\section{Conclusions and Future Work}

Identifying refactoring opportunities is a fundamental task for software development teams that aim to reduce their maintenance and evolution costs. Yet, it is still an open research problem. 

Recent papers have been successful in building models that learn from previous refactoring operations in open-source systems. 
In this paper, we experiment with such models at ING, a large financial company that is always interested in improving the quality of its code bases. 
After a series of empirical studies and observations, we conclude that machine learning models are able to predict future refactorings, with interesting levels of accuracy, also in industry systems. 

We see opportunities for future work. First, in the data collection process. The identification of methods that do not need refactoring is done through heuristics. ING (and we conjecture the same in other companies) does not have any labeled datasets of methods considered ideal from which models could learn from. Heuristics are therefore needed, and future work should explore different ones. Second, the models have much to improve. Our models achieve an accuracy of 74\% in unseen ING projects. While 74\% is an encouraging number, anecdotal evidence from industry suggests that tools should have at most a 15\% false positive rate. This means these models still need to improve before developers do not find them problematic. Finally, presenting these results to developers is still a challenge. At ING, we experimented with showing the predictions within our code review tools, but the engagement was low. While we have our own assumptions of why this was the case, building a developer-friendly tool is surely an important step before deploying refactoring recommendation models.

\begin{acks}

This project was partially funded by the EU ITEA3 
Industrial-grade Verification and Validation of Evolving Systems (IVVES) project, under grant agreement ``ITEA2019-18022-IVVES'', and by the ICAI AI for Fintech Research. 
We also thank the five ING experts who took their time to support our research.

\end{acks}

\bibliographystyle{ACM-Reference-Format}
\balance
\bibliography{library}


\end{document}